\documentclass{iopart}
\input{psfig}
\usepackage{epsfig}

\def\Mpc{\, h^{-1} \, {\rm Mpc}}

\def\Mpc{\ifmmode {\, h^{-1} \, {\rm Mpc}}
\else {$h^{-1}\,$ Mpc}\fi}

\def\s8{{\sigma_8}}

\def\ltsima{$\; \buildrel < \over \sim \;$}
\def\simlt{\lower.5ex\hbox{\ltsima}} 
\def\gtsima{$\; \buildrel > \over \sim \;$} 
\def\simgt{\lower.5ex\hbox{\gtsima}}

\def\omegam{{\Omega_{\rm m}}}
 
\def\omegab{{\Omega_{\rm b}}}
 
\def\omegal{{\Omega_\Lambda}}
 
\def\omegabh2{{\omegab h^2}} 

\begin{document}
\jl{6}
\title{Observational Tests of FRW World Models}
\author{
Ofer Lahav\footnote[1]{E-mail address: {\tt lahav@ast.cam.ac.uk}}}

\address{Institute of Astronomy, Madingley Road, Cambridge CB3 0HA,UK}
\date{}
\begin{abstract}
Observational tests for the Cosmological Principle  
are reviewed. Assuming the FRW metric we then summarize estimates 
of cosmological parameters from 
various data sets, in particular the Cosmic Microwave Background 
and the 2dF galaxy redshift survey.
These and other  analyses suggest a best-fit $\Lambda$-Cold Dark Matter
model with  
$\omegam = 1 -\omegal \approx  0.3$ and $H_0 \approx 70$ km/sec/Mpc.
It is remarkable that different measurements
converge to this `concordance model', although  
it remains to be seen if the two main components 
of this model, the dark matter and the dark energy, 
are real entities or just `epicycles'.
We point out some open questions related to this fashionable model.
\end{abstract}

\section{Introduction}

The Cosmological Principle (CP) was first adopted when observational
cosmology was in its infancy; it was then little more than a
conjecture, embodying 'Occam's razor' for the simplest possible
model. Observations could not then probe to significant redshifts, the
`dark matter' problem was not well-established and the Cosmic Microwave
Background (CMB) and the X-Ray Background (XRB)  were still unknown.  
If the  Cosmological Principle turned out to be invalid 
then the consequences to our understanding of cosmology would be dramatic, 
for example the conventional way of interpreting the age of the Universe, 
its geometry and matter content would have to be revised. 
Therefore it is 
important to revisit this underlying assumption in the light of new
galaxy surveys and measurements of the background radiations.

Like with any other idea about the physical world, we cannot 
prove a model, but only falsify it.
Proving  the homogeneity of the Universe is in particular difficult 
as we observe the Universe from one point in space, and we can only 
deduce isotropy directly.
The practical methodology we adopt is to assume homogeneity and to assess
the level of fluctuations relative to the mean, and hence to test
for consistency with  the underlying hypothesis.
If the assumption of homogeneity turns out to be wrong, then 
there are  numerous possibilities
 for inhomogeneous models, and each of them must be
tested against the observations.

Here we examine the degree of smoothness with scale 
by considering 
redshift and peculiar velocities surveys, 
radio-sources, the XRB, the Ly-$\alpha$ forest,
and the CMB.
We discuss some  inhomogeneous models
and show that a fractal model on large scales is highly improbable.
Assuming a
Friedmann-Robertson-Walker
(FRW) metric we evaluate the `best-fit Universe' by 
performing a joint analysis of cosmic probes.

\section{The Cosmological Principle(s)}

Cosmological Principles were stated over different periods in 
human history based on philosophical and aesthetic considerations 
rather than on fundamental physical laws.
Rudnicki (1995) summarized some of these principles in modern-day
language:

\bigskip

$\bullet$ The Ancient Indian: 
{\it The Universe is infinite in space and time and is 
infinitely heterogeneous}.

$\bullet$ The Ancient Greek:
{\it Our Earth is the natural centre of the Universe}.

$\bullet$ The Copernican CP:
{\it 
The Universe as 
observed from any planet looks much the same}.

$\bullet$ The Generalized CP:
{\it 
The Universe is (roughly) homogeneous and isotropic}.

$\bullet$ The Perfect CP:
{\it The Universe is (roughly) homogeneous in space and time,
and is isotropic in space}.

$\bullet$ The Anthropic Principle:
{\it A human being, as he/she is, can exist only in the Universe
as it is.}

\bigskip

We note that the Ancient Indian principle can be viewed as 
a `fractal model'. 
The Perfect CP led to the steady state model, 
which although more symmetric than the CP, 
was rejected on observational grounds.
The Anthropic Principle is  becoming popular again, e.g. in 
`explaining' a non-zero cosmological constant.
Our goal here is to quantify `roughly' in the definition of the 
generalized CP, and to assess if one may assume safely
the FRW metric of space-time.

\section {Probes of Smoothness}

\subsection {The CMB Temperature at High Redshift}  

The CMB provides  the strongest evidence for the CP.
Most of the observational tests for the CP 
really constrain isotropy rather than homogeneity,
as they are done from our point in space.
However, one can use atoms and molecules at high redshift 
as `hypothetical observers' that `report' to us the CMB 
temperature at that redshift. 
According to the standard Hot Big Bang 
model, which assumes the FRW metric, the CMB temperature
varies with redshift as  
$T(z) = T(0) (1+z)$, where $T(0) = 2.726 \pm 0.010$ K.
Recent measurements (Srianand et al. 2000)  of the relative populations
of atomic fine-structure levels, which are excited
by the background radiation, find from atoms
and molecules in an isolated cloud which happens to 
be along the line of sight to a quasar, $6 < T(z=2.34) < 14$ K, 
in very good agreement with the predicted temperature of 9.1 K.  
This provides yet another strong evidence for the Hot Big Bang,
together with the Hubble recession of galaxies 
and the Big Bang Nucleosynthesis.


\subsection {The CMB Fluctuations}

Ehlers, Garen and Sachs (1968) showed that by combining the 
CMB isotropy with the Copernican principle 
one can deduce homogeneity. More formally the 
EGS theorem (based on Liouville theorem) states that
``If the fundamental observers in a dust spacetime see an isotropic
radiation field, then the spacetime is locally FRW''.
The COBE measurements of temperature 
fluctuations  $\Delta T/T = 10^{-5} $ on scales of $10^\circ$ give 
via the Sachs Wolfe effect ($\Delta T/T = \frac {1}{3} \Delta \phi/c^2$) 
and Poisson equation
rms density fluctuations of ${{\delta \rho} \over {\rho}} \sim 10^{-4} $ on $1000 \Mpc$ (e.g. Wu,  Lahav \& Rees 
1999; see Fig 3 here), i.e. the deviations from a smooth Universe are tiny.

\subsection {Galaxy Redshift Surveys}

 The distribution of galaxies in local 
 redshift surveys (e.g. ORS and IRAS) is highly 
 clumpy, with the Supergalactic Plane seen in full glory.
 However, deeper surveys such as LCRS and 2dFGRS (Figure 1) 
 show that the fluctuations 
 decline as the length-scales increase. Peebles (1993) has shown 
 that the angular correlation functions for the Lick and APM surveys 
 scale with magnitude as expected in a universe which approaches 
 homogeneity on large scales.

\begin{figure}
\protect\centerline{
\psfig{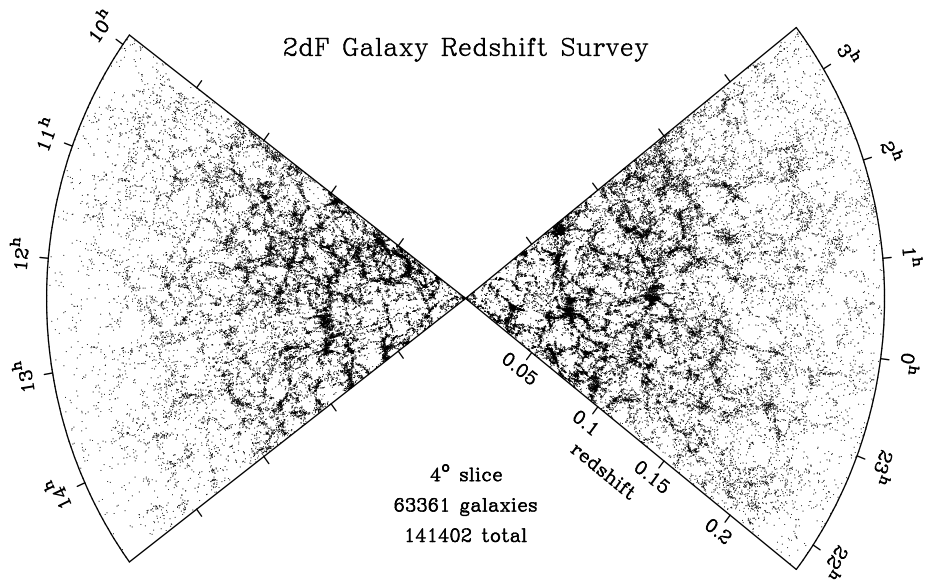}}
\caption[]{
The distribution of over 63,000 galaxies in the 2dFGRS, 
drawn from a total of 210,00 galaxies. 
The slices are 4 deg thick, towards the Northern Galactic Pole (left) 
and towards the Southern Galactic Pole (right). 
Not all 2dF fields within the slice have been observed at this stage; 
hence there are weak variations of the density of sampling. 
The image reveals a wealth of structure, 
including superclusters and voids,
but the similarity between the two slices suggests that on large scales
the universe is isotropic and homogeneous 
(from Peacock et al. 2001).  }
\end{figure}

Multifibre technology now allows us to measure  redshifts
of millions of galaxies.  Two major surveys are underway.
The US Sloan Digital Sky Survey (SDSS) will measure redshifts to about
1 million galaxies over a quarter of the sky.  The Anglo-Australian 2
degree Field Galaxy Redshift Survey (2dFGRS) survey 
has already 
measured redshifts for 210,000 galaxies
selected from the APM catalogue.  
(as of November 2001).  The median redshift of both the
SDSS and the 2dFGRS galaxy redshift surveys is ${\bar z} \sim 0.1$.  While
they can provide interesting estimates of the fluctuations on 
intermediate scales (e.g. Peacock et al. 2001; Percival et al. 2001; 
see Fig 2 here),
the problems of biasing, evolution and
$K$-correction, would limit the ability of SDSS and 2dF to `prove' the
Cosmological Principle.  (cf. the analysis of the ESO slice by
Scaramella et al 1998 and Joyce et al. 1999).
But the  measurement of the power spectrum 
of 2dFGRS (Percival etal. 2001) shows good agreement with 
what is expected in both shape and amplitude from  
the $\Lambda$+ Cold Dark Matter (CDM) model
(which assumes an underlying FRW metric).

Assuming a Gaussian prior on the Hubble constant $h=0.7\pm0.07$ (based
on Freedman et al. 2000) the shape of the recovered power spectrum
within $0.02 <k <0.15 h$/Mpc (Figure 2) was used to yield 68\%
confidence limits the shape parameter $\omegam h=0.20 \pm 0.03$, and
the baryon fraction $\omegab/\omegam=0.15 \pm 0.07$, in accordance
with the popular `concordance' model
\footnote{As shown in Percival et al. 2001 , the likelihood analysis
gives a second (non-standard) solution, 
with   $\omegam h \sim 0.6$, and
the baryon fraction $\omegab/\omegam=0.4$,
which generates baryonic `wiggles'.
The `wiggles' are probably due to `noise' 
correlated by the survey window function.
},
in particular as the derived 
from the CMB data (Efstathiou et al. 2001; Lahav et al. 2001).

\subsection{Peculiar Velocities}

Peculiar velocities are powerful as they probe directly the mass distribution
(e.g. Dekel et al. 1999).
Unfortunately, as distance measurements increase with distance, 
the scales probed are smaller than the interesting 
scale of transition to homogeneity.
Conflicting results 
on both the amplitude  and coherence of the flow suggest that
peculiar velocities cannot yet set strong constraints on the amplitude
of fluctuations on scales of hundreds of Mpc's.
Perhaps the most promising method for the future is the 
kinematic Sunyaev-Zeldovich effect which allows one to measure the 
peculiar velocities of clusters out to high redshift.

The agreement between the CMB dipole and the dipole anisotropy 
of relatively nearby galaxies argues in favour of large scale 
homogeneity.
The IRAS dipole (Strauss et al 1992, Webster et al 1998, 
Schmoldt et al 1999)
shows an apparent convergence 
of the dipole, with 
misalignment angle of only $15^\circ$.
Schmoldt et al. (1999) claim that 
2/3  of the dipole arises from within a $40 \Mpc$, 
but again it is difficult
to `prove' convergence from catalogues of finite depth.

 \subsection{Radio Sources}

Radio sources in surveys have typical median redshift
${\bar z} \sim 1$, and hence are useful probes of clustering at high
redshift. 
Unfortunately, it is difficult to obtain distance information from
these surveys: the radio luminosity function is very broad, and it is
difficult to measure optical redshifts of distant radio sources.
Earlier studies
claimed that  the distribution of radio sources supports the 
`Cosmological Principle'.
However, 
the wide range in intrinsic luminosities of radio sources
would dilute any clustering when projected on the sky.  
Recent analyses  of
new deep radio surveys (e.g. FIRST)
suggest that radio sources are actually  clustered at least as strongly
as local optical
galaxies 
(e.g. Cress et al. 1996; Magliocchetti et al. 1998).
Nevertheless, on the very large scales the distribution of radio sources
seems nearly isotropic. 
Comparison of the measured dipole, quadrupole and higher harmonics
in a radio sample 
in the Green Bank and Parkes-MIT-NRAO
4.85 GHz surveys 
to the theoretically predicted ones (Baleisis et al. 1998)
offers a crude upper limits of the fluctuations on scales $ \sim 600
h^{-1}$ Mpc, consistent with 
the $\Lambda$-CDM model.

\begin{figure}
\protect\centerline{
\psfig{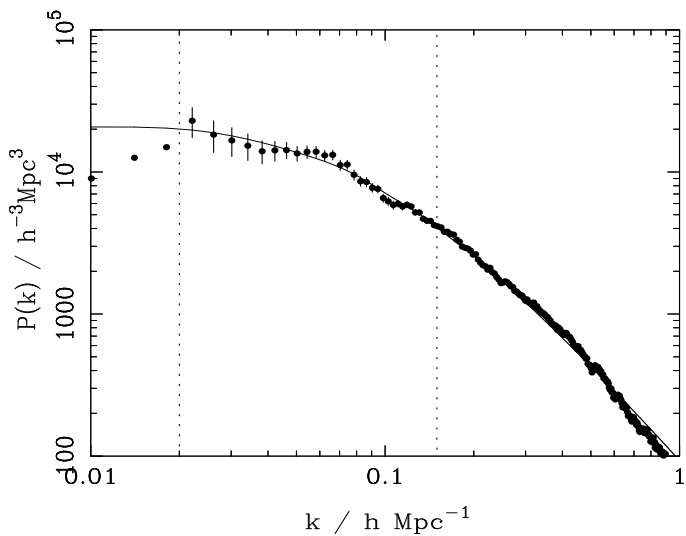}}
\caption[]{The observed (convolved) 2dFGRS power-spectrum
(Percival et al. 2001), 
and a linear theory $\Lambda$-CDM 
(real space, convolved with the window function)
with $\omegam h = 0.2, \omegab/\omegam = 0.15, h=1, n=1$ 
and best-fitting (linear) redshift space 
normalization 
$\sigma^S_{8g} = 0.94$.
Only the linear regime $ 0.02 < k < 0.15 \;h/$Mpc was used 
to derive the above parameters 
(roughly corresponding to CMB harmonics $200 <  l < 1500$
for a flat $\omegam = 0.3$ universe).} 
\label{2dF_Pk}
\end{figure}

 \subsection {The XRB}

 Although discovered in 1962, the origin of
 the X-ray Background (XRB) is still unknown,  
 but is likely
 to be due to sources at high redshift 
 (for review see Boldt 1987; Fabian \& Barcons 1992).
 The XRB sources are probably
 located at redshift $z < 5$, making them convenient tracers of the mass
 distribution on scales intermediate between those in the CMB as probed
 by COBE, and those probed by optical and IRAS redshift
 surveys.

The interpretation of the results depends somewhat on the nature of
the X-ray sources and their evolution.  
By comparing
the predicted multipoles to those observed by HEAO1 
(Lahav et al. 1997; Treyer et al. 1998; Scharf et al. 2000)
we estimate the amplitude of fluctuations for an
assumed shape of the density fluctuations 
(e.g. CDM models).  
The observed fluctuations in the XRB
are roughly as expected from interpolating between the
local galaxy surveys and the COBE and other CMB experiments.
The rms fluctuations 
${ {\delta \rho} \over {\rho} }$
on a scale of $\sim 600 h^{-1}$Mpc 
are less than 0.2 \%.

\subsection {The Lyman-$\alpha$ Forest}


The Lyman-$\alpha$ 
forest reflects the neutral hydrogen distribution and therefore
is likely to be a more direct  trace of the mass distribution 
than galaxies are.
Unlike galaxy surveys which are
limited to the low redshift Universe, the forest spans a large
redshift interval, typically $1.8 < z < 4$, corresponding 
to comoving interval of $\sim 600 \Mpc$.
Also, observations of the
forest are not contaminated by complex selection effects such as those
inherent in galaxy surveys.  It has been suggested qualitatively by
Davis (1997) that the absence of big voids in the distribution of 
Lyman-$\alpha$
absorbers is inconsistent with the fractal model.
Furthermore, all lines-of-sight towards quasars look
statistically similar.  
Nusser \& Lahav (2000) 
predicted the distribution of the flux  in Lyman-$\alpha$ 
 observations in a specific
truncated fractal-like model. They found that indeed in this model there are
too many voids compared with the observations and conventional (CDM-like)
models
for structure formation.
This too supports the common view that on large scales the Universe 
is homogeneous.

\subsection {The Isotropy of the Distribution of SN Ia} 

Another test for isotropy, based on the distribution of 79
Supernovae Ia
out to redshift $z \approx 1$ is described in Kolatt \& Lahav (2001).
They divided the sky into two hemispheres that give the most 
{\it discrepant} values of $\omegam$ and $\omegal$. 
For a perfect FRW Universe, Monte Carlo realizations
that mimic the observed set of SN Ia, yield 
values higher than the measured discrepancy in about 20 \% of the case.
It would be interesting to repeat this  
isotropy test with future SN Ia experiments, e.g. SNAP.

\section {Is the Universe a Fractal ?}

The question of whether the Universe 
is isotropic and homogeneous on large scales
can also be  phrased in terms of the fractal structure of the 
Universe.
A fractal is a geometric shape that is not homogeneous, 
yet preserves the property that each part is a reduced-scale
version of the whole.
If the matter in the Universe were actually 
distributed like a pure fractal on all scales then the 
Cosmological Principle 
would be invalid, and the standard model in trouble.
Current data already strongly constrain any non-uniformities in the 
galaxy distribution (as well as the overall mass distribution) 
on scales $> 300 \Mpc$.

If we count, for each galaxy,
the number of galaxies within a distance $R$ from it, and call the
average number obtained $N(<R)$, then the distribution is said to be a
fractal of correlation dimension $D_2$ 
if $N(<R)\propto R^{D_2}$. Of course $D_2$
may be 3, in which case the distribution is homogeneous rather than
fractal.  In the pure fractal model this power law holds for all
scales of $R$.

The fractal proponents (Pietronero et al. 1997)  have
estimated $D_2\approx 2$ for all scales up to $\sim 500\Mpc$, whereas
other
groups 
have obtained scale-dependent values 
(for review see Wu et al. 1999 and references therein).

Estimates of $D_2$ 
from the CMB and the XRB
are consistent with $D_2=3$ to within
$10^{-4}$  on the very
large scales (Peebles 1993; Wu et al. 1999).
While we reject the pure fractal model in this review, the performance
of CDM-like models of fluctuations on large scales have yet to be
tested without assuming homogeneity {\it a priori}. On scales below,
say, $30 \Mpc$, the fractal nature of clustering implies that one has
to exercise caution when using statistical methods which assume
homogeneity (e.g. in deriving cosmological parameters).  
We emphasize that we only considered
one `alternative' here, which is the pure fractal model where $D_2$ is a
constant on all scales.

\section {More Realistic Inhomogeneous Models} 

As the Universe appears clumpy on small scales it is clear that 
assuming the Cosmological Principle and the FRW metric is only an 
approximation, and one has to 
average carefully the density in  
Newtonian Cosmology (Buchert \& Ehlers 1997).
Several models in which the matter in clumpy 
(e.g. 'Swiss cheese' and voids)
have been proposed 
(e.g. Zeldovich 1964; Krasinski 1997; Kantowski 1998; Dyer \& Roeder  
1973; Holz \& Wald 1998;  C\'el\'erier 1999; Tomita 2001). 
For example, if the line-of-sight to a distant  object is `empty' 
it results in a gravitational lensing de-magnification of the object.
This modifies the  FRW luminosity-distance relation, with 
a clumping factor as  another free parameter. 
When applied to a sample of SNe Ia
the density parameter  of the Universe
$\omegam$ could be underestimated if FRW is used 
(Kantowski 1998; Perlmutter et al. 1999).
Metcalf \& Silk (1999) pointed out that this effect can be used as a test 
for the nature of the dark matter, i.e. to test if it is smooth  
or clumpy.

\section {Cosmological Parameters from a Joint Analysis}

\subsection {A Cosmic Harmony ? }

Different cosmic probes determine different sets of cosmological parameters.
Below we give the approximated dependence on the parameters
$\omegam, \omegab, \omegal, h$ and the (linear theory) 
normalization $\sigma_{8{\rm m}}$ of the mass fluctuations in $8 \Mpc$ spheres.

\begin{itemize}

\item CMB: 
$\omegal + \omegam, \;
\Omega_m h^2, \;\Omega_b h^2,\;   
\sigma_8$

\item SNIa:  
$3 \omegal - 4 \omegam$

\item Redshift surveys:  
$\omegam^{0.6}/b, \; b \sigma_8, \;\omegam h$

\item Peculiar velocities:
$\sigma_8 \omegam^{0.6}, \;\omegam h$

\item Cluster abundance:
 $\sigma_8 \omegam^{0.6}$

\item Weak lensing:
$\sigma_8 \omegam^{0.6}$

\item Baryon fraction:
$\omegab, \; \omegam, \; h$

\item Cepheids, SZ, time-delay: $h$

\end{itemize} 

The dependence on  the factor of roughly $\omegam^{0.6}$ in different probes
(peculiar velocities, cluster abundance and cosmic shear) 
is a coincidence.
The important point is that by using `orthogonal' constraints one can 
significantly improve the estimation of cosmological parameters.
By performing joint likelihood analyses, one can
overcome intrinsic degeneracies inherent in any single analysis
and so estimate fundamental
parameters much more accurately. The comparison of
constraints can also provide a test for the validity of the assumed
cosmological model or, alternatively, a revised evaluation of the
systematic errors in one or all of the data sets.  Recent papers that
combine information from several data sets simultaneously include
Webster et al. (1998); Lineweaver (1998); 
Gawiser \& Silk (1998),  Bridle et al. (1999, 2001), 
Eisenstein, Hu \& Tegmark (1999); Efstathiou et al. (1999, 2001);  
Bahcall etal. (1999) and  Wang et al. (2000).

\subsection {Statistical Issues }

While joint Likelihood analyses employing both CMB and LSS data are
allowing more accurate estimates of cosmological
parameters, they involve various subtle statistical issues:
\begin{itemize}

\item There is the uncertainty that a sample does not represent 
a typical patch of the FRW Universe to yield reliable global cosmological 
parameters.
\item The choice of the model parameter space is somewhat arbitrary.
\item One commonly solves for the probability for the data given a model
      (e.g. using a Likelihood function),  
      while in the Bayesian framework this should be modified
      by the prior for the model and its parameters.
\item If one is interested in a small set of parameters, should one marginalise
      over all the remaining parameters, rather than  fix them at certain 
      (somewhat ad-hoc) values ?  
\item The `topology' of the Likelihood contours may not be simple. 
      It is helpful when the Likelihood contours of different probes 
      `cross' each other to yield a global maximum 
       (e.g. in the case of CMB and SNe), but in other cases
       they may yield distinct separate `mountains', and the joint 
       maximum Likelihood may lie in a `valley'.
\item Different probes might be spatially correlated, i.e. 
       not necessarily independent.
\item What weight should one give to each data set ?
\end{itemize}

In a long-term collaboration in Cambridge (Bridle
et al. 1999, 2001; Efstathiou et al. 1999; Lahav et al. 2000) we have
compared and combined in a self-consistent way various
cosmic probes: CMB, galaxy redshift surveys, galaxy cluster number
counts, type Ia Supernovae and galaxy peculiar velocities.  These
analyses illustrate the power of combining different data sets for
constraining the fundamental parameters of the Universe.  Our analysis
suggests, in agreement with studies by other groups
(e.g. Bahcall et al. 1999, Wang et al. 2000), that we live in a
flat accelerating Universe, 
with about 30\% of the critical density in the
form of matter (baryonic and non-baryonic) and 70\% in the form of `dark
energy' (which may be Einstein's cosmological constant, or a
 more complicated dynamical vacuum energy). 
We have also addressed recently (Lahav et al. 2000; Lahav 2001) the issue of
combining different data sets, which may suffer different systematic
and random errors.  We generalised the standard procedure of combining
likelihood functions by allowing freedom in the relative weights of
various probes.  This is done by including in the joint likelihood
function a set of `Hyper-Parameters', which are dealt with using
Bayesian considerations.  The resulting algorithm, which assumes
uniform priors on the logarithm of the Hyper-Parameters, is simple to
implement.

\section{The Best-Fit Concordance Model}

Although the $\Lambda$-CDM model
with comparable amounts of dark matter and dark energy is rather esoteric,
it is remarkable that different measurements
converge to `concordance model' with parameters: 

\begin{itemize}

\item $\Omega_k  \approx 0$, 

\item $\Omega_m =1 - \Omega_\Lambda \approx 0.3$, 

\item $\Omega_b h^2 \approx 0.02$,
 
\item $h \approx 0.7$

\item the age of the Universe $t_0 \approx 14 $ Gyr, 

\item  the spectral index $n \approx 1$,

\item $\sigma_{8{\rm m}} \approx 0.7$.

See for example Figures 2 and 3, that show that roughly the same
parameters fit well the 2dFGRS power spectrum and the CMB fluctuations .
Perhaps the least accurate estimates on that list are for $\omegam$ 
and $\sigma_{8{\rm m}}$.

\end{itemize}

\bigskip

\begin{figure}
\protect\centerline{
\epsfig{file=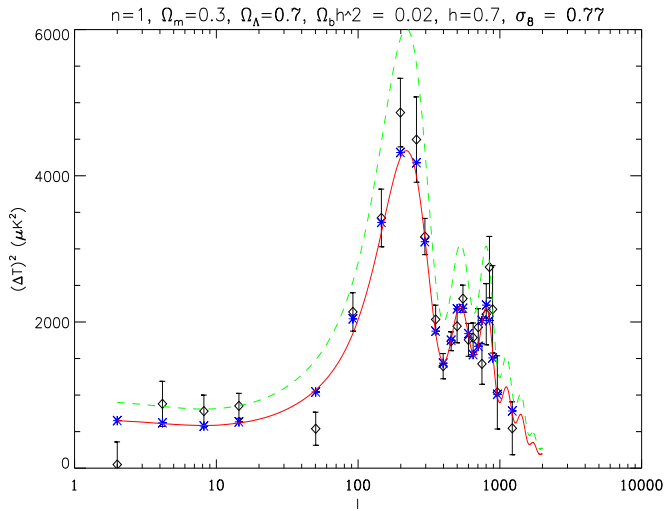,width=9cm,height=7cm}}
\caption[]{A compilation of the latest CMB 
data points 
$T_0^2 \ell(\ell +1) C_{\ell}/(2 \pi)$
(open circles with error bars, from WTZ01)
against spherical harmonic $\ell$.  The line
shows the predicted angular power-spectrum for a $\Lambda$-CDM model
with $n=1$, $\Omega_{\rm m} = 1 -\Omega_\Lambda = 0.3$, $\Omega_{\rm
b} h^2 = 0.02$ (BBN value), $h=0.7$, 
COBE normalization 
$\sigma_{8m} =0.90$ (dashed line) and the lower best-fit to WTZ01 data 
$\sigma_{8m} =0.77$ (solid line). 
Recent results from cluster abundance (e.g. Seljak 2001) 
are actually in agreement with this low normalisation.
The stars indicate this best fit model convolved with the 
experimental window functions.
A similar model is also
the best-fit to the shape of the 2dF galaxy power-spectrum (Figure 2).}
\label{dt}
\end{figure}

\section{Discussion}

Analysis of the CMB, the XRB, radio sources and the Lyman-$\alpha$ 
which probe scales of 
$\sim 100-1000 \Mpc$ strongly support the Cosmological Principle
of homogeneity and isotropy.
They rule out  a pure fractal model.
However, there is a need for more realistic  inhomogeneous models
for the  small scales. This is in particular important for
understanding the validity of cosmological parameters obtained
within the standard FRW cosmology.

While phenomenologically the $\Lambda$-CDM
model has been successful in fitting a wide range
of cosmological data, there are some open questions:

\begin{itemize}

\item
Both components of the model, $\Lambda$ and CDM, 
have not been directly measured.
Are they `real' entities or just `epicycles' ?

\item
Why is  $\Omega_m \sim \Omega_\Lambda$ at the present-epoch ?
Do we need to introduce  a new physics 
or invoke the Anthropic Principle to explain it ?

\item
There are still open problems in 
$\Lambda$-CDM on the small scales
e.g. galaxy profiles and satellites
(e.g. Selwood \& Kosowsky 2000).

\item 
The age of the Universe is uncomfortably close to some estimates for the 
age of the Globular Clusters, when  their epoch of 
formation is also taken into account (Gnedin, Lahav \& Rees 2001). 

\item  
Could other (yet unknown) models fit the data equally well ?

\item
Where does the field go from here ?
Would the activity focus on refinement of 
the cosmological parameters within $\Lambda$-CDM, 
or on introducing entirely new paradigms   ?

\end{itemize} 

These issues will no doubt 
be revisited soon with larger and more accurate data sets.
We will soon be able to map
the fluctuations with scale and epoch, and to analyze jointly
redshift surveys 
(2dF, SDSS) and
CMB (MAP, Planck) data. 
These high quality data sets 
will allow us to study a wider range of models and parameters.

\section*{Acknowledgments}

I thank my collaborators for their contribution to the work
presented here, and to  the  oraganisers 
for the hospitality in Cape Town.

\bigskip



\end{document}